\documentclass[aps,prl,notitlepage,reprint,floatfix,amsmath,amssymb]{revtex4-2}

\usepackage{graphicx}
\usepackage{dcolumn}
\usepackage{bm}

\usepackage[separate-uncertainty = true]{siunitx}
\usepackage[normalem]{ulem}
\usepackage{soul}

\begin{document}


\title{Observation of quasi bound states in open quantum wells of cesiated p-doped GaN surfaces}
\author{Myl\`{e}ne Sauty$^{1,2}$}
\author{Jean-Philippe Banon$^{1,3,4}$}
\author{Nicolas M. S. Lopes$^{1}$}
\author{Tanay Tak$^{5}$}
\author{James S. Speck$^{5}$}
\author{Claude Weisbuch$^{1,5}$}
\author{Jacques Peretti$^{1}$}
\affiliation{$^1$Laboratoire de Physique de la Mati\`{e}re Condens\'{e}e, CNRS, Ecole polytechnique, Institut Polytechnique de Paris, 91120 Palaiseau, France}
\affiliation{$^2$Service de Physique de l’État condensé (SPEC), Université Paris-Saclay, CEA-CNRS, F-91191 Gif-sur-Yvette, France}
\affiliation{$^3$Laboratoire Charles Fabry, Institut d’Optique Graduate School, CNRS, Université Paris-Saclay, 91127 Palaiseau, France}
\affiliation{$^4$Université Jean Monnet Saint-Etienne, CNRS, Institut d’Optique Graduate School, Laboratoire Hubert Curien, UMR 5516, F-42023 Saint-Etienne, France}
\affiliation{$^5$Materials Department, University of California, Santa Barbara, California 93106, USA}

\date{\today}

\begin{abstract}
The electron density of states in the open quantum well formed by the downward band bending region at the surface of cesiated \textit{p}-type GaN is investigated. We theoretically predict the existence of metastable resonant states in this non confining potential with an intrinsic lifetime around 20~fs. Their experimental observation requires access to the empty conduction band of the cesiated semiconductor, which is possible with near-band gap photoemission spectroscopy. The energy distribution of the photoemitted electrons shows contributions coming from electrons accumulated into the resonant states at energies which agree with calculations. 
\end{abstract}

\maketitle

The quantized energy states in narrow, triangular-shaped space-charged layers formed at metal-insulator-semiconductor (MIS) interfaces and their influence on carrier mobilities and transport properties have been investigated for more than 50 years~\cite{Schrieffer1957, Ando1982, Stern1973}. Under specific band alignments, similar triangular quantum wells form at semiconductor-semiconductor interfaces, as in modulation-doped (MD) AlGaAs/GaAs or AlGaN/GaN interfaces, leading to high mobility materials, novel phenomena such as the quantum Hall effect (although first observed in Si MOS structures) and electronic devices such as high electron mobility transistors~\cite{Dingle1978, Jones2016}. Such narrow triangular wells for carriers can also form at the free surface of semiconductors depending on doping and surface states \cite{Ando1982, Kawaji1968}. Typically, the clean surface of \textit{p}-doped GaAs or GaN displays a downward band bending region (BBR) due to the mid gap pinning of the Fermi level at the surface~\cite{Scheer1965}.

The energy position and properties of these quantized states have been particularly studied when the well is closed and contains carriers, forming a two-dimensional electron or hole gas (2DEG or 2DHG) \cite{Ando1982, Stern1973, Dingle1978, Jones2016}, in MIS inversion layers or MD structures, but also at the free surface of semiconductors with a a high electron affinity and a specific pinning of the Fermi level. 

Optical techniques were used to measure photoluminescence from the 2DEG formed at the AlGaAs/GaAs and AlGaN/GaN interfaces~\cite{Bergman1991, Bergman1996, Kaneriya2023}, as well as the 2DHG at the GaN/AlGaN interface~\cite{Mechin2024}.
Electrical measurements, particularly magnetotransport measurements of Shubnikov–de Haas oscillations, gave access to the energy separation between the quantized states and to the carrier concentrations in each energy level, in AlGaN/GaN~\cite{Zheng2000}, or at the free surface of Te~\cite{Bouat1978}. 
Finally, photoemission spectroscopy allowed the direct measurement of the energy position, momentum dispersion and carrier dynamics in the quantized states of surface inversion layers, typically in InAs, InSb or InSe where the pinning of the Fermi level could be tuned with surface impurities, dopants or alkali metals~\cite{Olsson1996, Aristov1999, Wutke2023, Chen2020}. 

The energy landscape for carriers around the surface BBR of a semiconductor can, however, be significantly different from the case of inversion layers, as for example at the free surface of a p-doped semiconductor after the deposition of a monolayer of an alkali metal, a widely used technique for photocathodes~\cite{Scheer1965, Spicer1977, Wang2021}. In this case, the semiconductor workfunction is reduced such that the vacuum level is below the bulk conduction band minimum (CBM), to the situation of negative electron affinity (NEA), while the Fermi level is pinned at mid-gap, well below the CBM. In this configuration, the triangular well formed by the surface BBR is nonconfining on the vacuum side, hereafter referred to as the open quantum well. 
In this work, we show that although this potential profile induces a continuum of states in the BBR, there still exist resonant states manifested by local maxima in the density of states (DOS). The experimental observation of these resonant states requires us to probe the empty conduction band of the semiconductor, contrary to the investigation of inversion layers described previously. 
There have been some attempts to observe such states using low energy photoemission spectroscopy on GaAs photocathodes \cite{Orlov2000, Korotkikh1978, Jin2015}. However, their contributions to the observed photoemission current could not be distinguished from those originating from the bulk states of the semiconductor \cite{Drouhin1985} because of the narrow band gap of GaAs and the use of an above band gap excitation energy, as will be discussed later.
In this work, we investigate instead the resonant states in the BBR of cesiated \textit{p}-GaN, a semiconductor with a much larger band gap than GaAs. The \textit{p}-doping of GaN induces a downward BBR of an amplitude of more than 1.5~eV over a few nanometers only, leading to quantization energies of the order of the electronvolt, much larger than in GaAs. We observe the signature of the resonant states in the energy distribution of the photoemitted electrons. Below band gap excitation allows us to clearly distinguish them from the bulk contribution.

\begin{figure*}[]
\centering
\includegraphics[width = 0.32\linewidth, trim=0cm 0cm 0cm 0cm, clip]{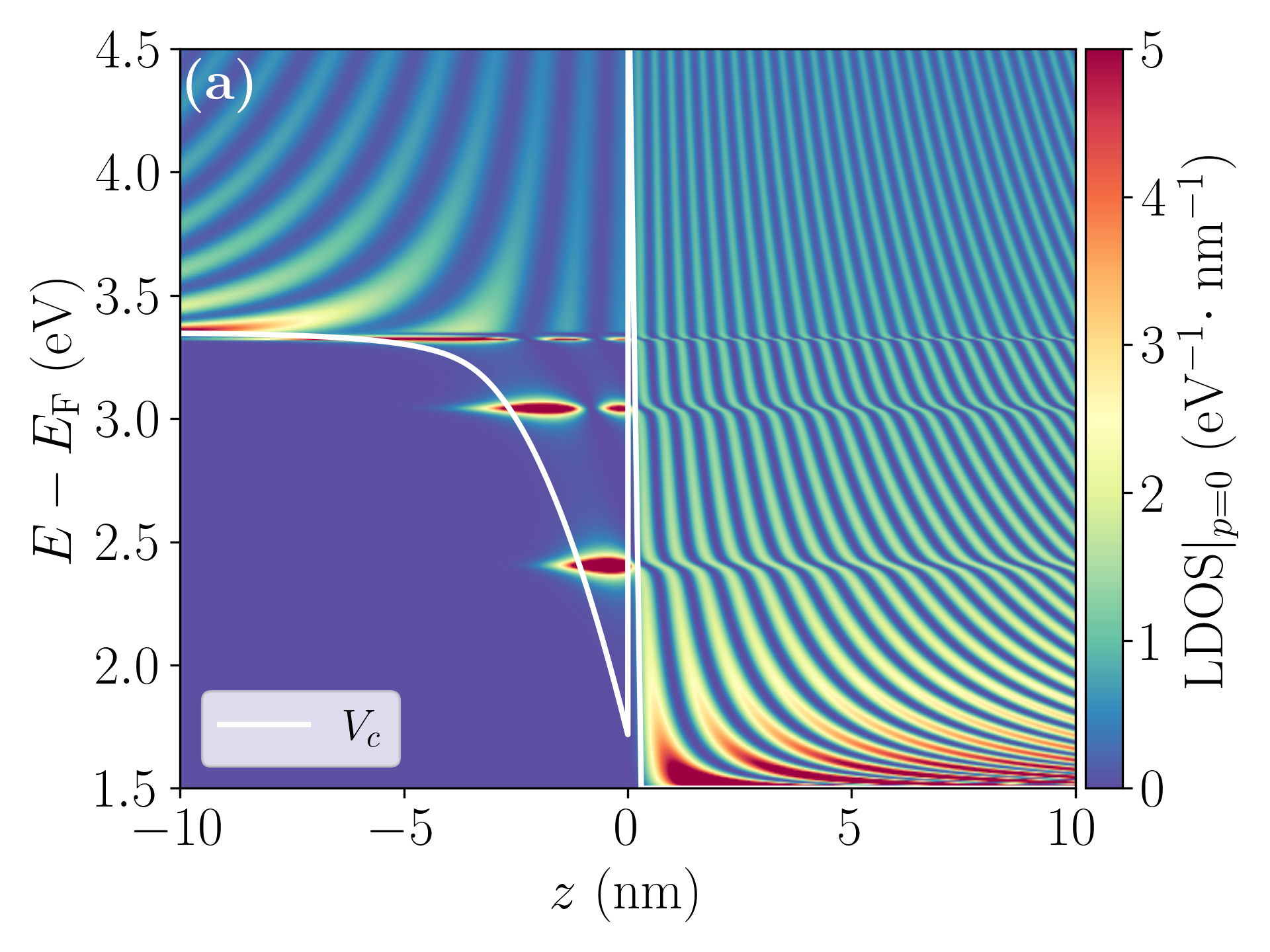}
\includegraphics[width = 0.32\linewidth, trim=0cm 0cm 0cm 0cm, clip]{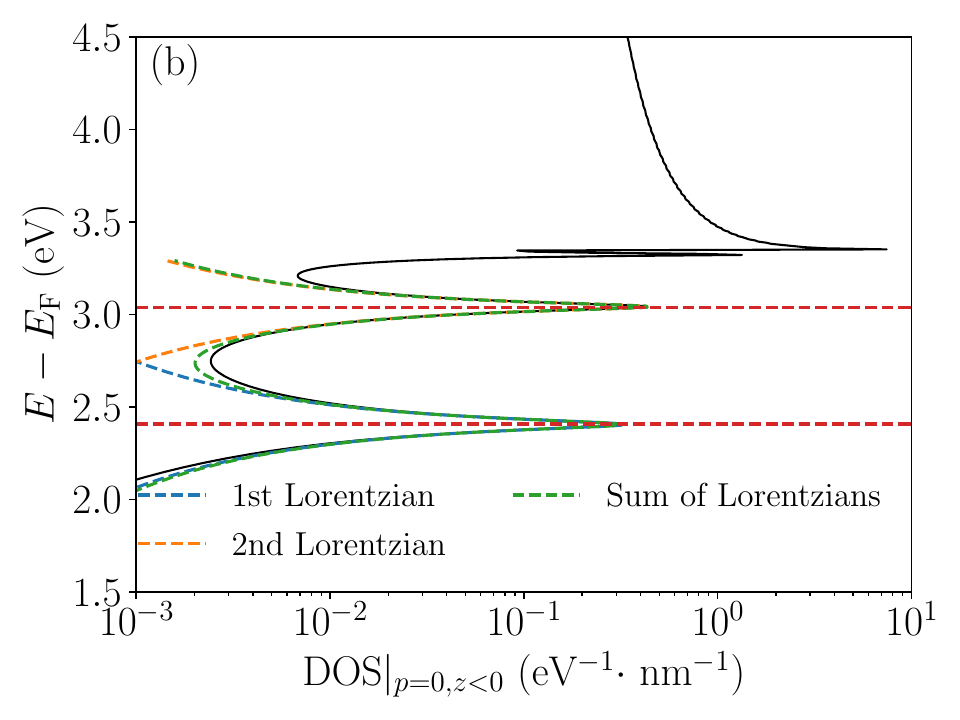}
\includegraphics[width = 0.32\linewidth, trim=0cm 0cm 0cm 0cm, clip]{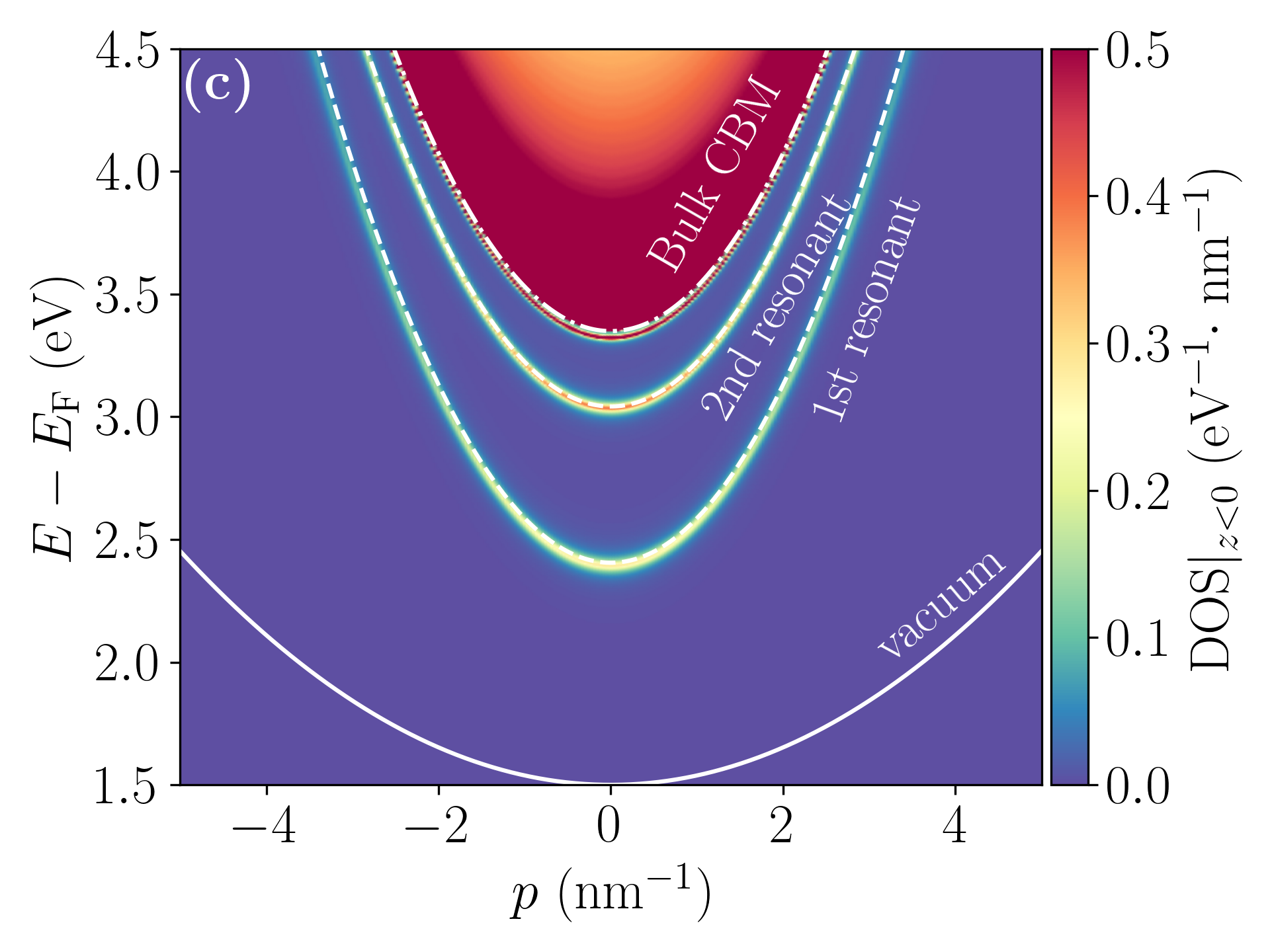}
\caption{\label{fig:DOS} (a) One-dimensional LDOS at vanishing transverse wave-vector ($p=0$) calculated with the envelope function approximation. The potential seen by electrons is marked by the white line, and corresponds to the CBM in the semiconductor and to the vacuum level $E_\mathrm{vac}$ in the vacuum. The LDOS shows that electron wavefunctions have a planewave structure both in vacuum for all energies and in the semiconductor for energies higher than the bulk CBM. For energies lower than the bulk CBM in the BBR, there is still a continuum of states but two resonant states appear at 2.4 and 3.0~eV. (b) 1D DOS at $p=0$, obtained by integration of the LDOS over $z <0$. The DOS is fitted with the sum of two Lorentzian functions of the form $A_1/[(E-E_1)^2 +(\Gamma_1 /2)^2] + A_2/[(E-E_2)^2 +(\Gamma_2 /2)^2]$. The fitted energy and spectral width of the resonances are $E_1 = 2.4$~eV, $E_2 = 3.0$~eV, $\Gamma_1 = 37.7$~meV and $\Gamma_2 = 28.2$~meV. The corresponding lifetimes are $\tau_1 = \hbar /\Gamma_1 = 17.5$~fs and $\tau_2 = 23.4$~fs.
 (c) 1D DOS conditioned to $z < 0$ as a function of the transverse wavevector $p$ (i.e. $\mathrm{DOS}|_{z<0} (p,E) = \int_{-L}^0 \mathrm{LDOS}(p,z,E) \, \mathrm{d}z / L$). The solid line corresponds to the dispersion relation in vacuum ($E = E_\mathrm{vac} + \hbar^2 p^2 / 2 m_0$). The dashed lines correspond to approximated dispersion relations for the two resonant states ($E \approx E_n + \hbar^2 p^2 / 2 m_e$) and the dashed dotted line corresponds to that of the bulk CBM $E = V_c(-\infty) + \hbar^2 p^2 / 2 m_e$.} 
\end{figure*}

Theoretically, the question of the existence and properties of quantized states in the surface BBR of cesiated GaAs photocathodes was addressed early on using analytical effective mass calculations~\cite{Korotkikh1978, Gerchikov1996}.  However, a vanishing envelope wavefunction was considered at the interface, as for infinite quantum wells, not really accounting for the nonconfining nature of the BBR well on the vacuum side. More recently, authors of numerical works on photoemission have used the effective mass model with a jump of effective mass at the interface, to find a transmission coefficient to be implemented in Spicer's three-step model~\cite{Zhuravlev:2014, Alperovich2021}, but we are not aware of a subsequent study of the metastable quantized states induced by this open quantum well. 

Here, we numerically investigate the local DOS (LDOS) in the open well at the free surface of cesiated \textit{p}-doped GaN. Resonances can be predicted by determining the local maxima of the DOS~\cite{Moiseyev:2011}. 
The calculation is achieved within an open system framework in which the electronic states in the semiconductor are represented in the envelope function approximation and are coupled to the plane wave states in the vacuum~(see Sec.~I in the Supplemental Material (SM) \footnote{See Supplemental Material in the ancillary section on arXiv for details on the theoretical model and corresponding numerics, as well as for a discussion of previous attempts to observe experimentally BBR resonant states in the literature, which also include Refs.~\cite{Orlov2000,Korotkikh1978,Jin2015,Drouhin1985, Zhuravlev:2014,Akkermans_Montambaux:book,Weisbuch:book, Bastard:1981,Levy-Leblond:1995,Einevoll:1994,BornWolf,singh_2003,nr}} for more details). The potential experienced by the conduction band electrons $V_c$ corresponds to the CBM in the semiconductor and to the vacuum level position (here 1.5~eV above the Fermi level $E_\mathrm{F}$) in vacuum. 
The Cs layer is modeled by a triangular potential barrier of height 5~eV, corresponding to the GaN workfunction before cesiation, and a thickness of 0.3~nm \cite{Sauty-LEEM-Cs, Wang2021} \footnote{Note that, a change in the values of the barrier height and thickness within a range of an eV and a few $\SI{}{\AA}$, respectively, does not affect significantly the simulation results}.
The spatial variation of the potential in the semiconductor close to the surface was computed by solving the Poisson equation assuming a Mg doping with concentration of $\SI{1e20}{\cm^{-3}}$, consistent with the experimental sample studied below, while imposing the Fermi level to be pinned at mid gap at the surface~(Sec.~I.C in the SM). As shown by the white solid line in Figure~\ref{fig:DOS}(a), this potential does not strictly form a well for electrons which can tunnel through the thin potential barrier at the surface. For such open systems, it is convenient to work with the Green's function $G$ associated to the effective mass Schr\"{o}dinger equation, defined by (Sec.~I.B in the SM)
\begin{equation}
    \frac{\hbar^2 }{2} \nabla \cdot \left[ \frac{ \nabla G }{m(z)} \right](\mathbf{r},\mathbf{r}^\prime, E) + ( E - V_c(z)) G (\mathbf{r},\mathbf{r}^\prime, E) = \delta(\mathbf{r} - \mathbf{r}^\prime) \: ,
    \label{eq:Sch:Green}
\end{equation}
and subjected to outgoing wave radiation conditions at infinity. Here $E$ denotes the energy variable, and $\mathbf{r}$ and $\mathbf{r}^\prime$ may be viewed as an observation and a source point, respectively.
The system being invariant by translations in the $(x,y)$ plane, the Green's function may be expressed as $G(\mathbf{r}_\parallel - \mathbf{r}_\parallel^\prime, z, z^\prime,E)$ where $\mathbf{r}_\parallel = (x,y)$ denotes the projection of a point $\mathbf{r}$ in the $(x,y)$-plane. The two-dimensional Fourier transform of the Green's function in the $(x,y)$ plane, $\bar{G}(\mathbf{p},z,z^\prime,E)$ then satisfies the equation
\begin{align}
        &\frac{\hbar^2 }{2} \frac{\mathrm{d}}{\mathrm{d}z} \left[ \frac{1}{m(z)}  \frac{\mathrm{d} \bar{G}}{\mathrm{d}z}  \right] (\mathbf{p},z,z^\prime,E) \nonumber\\
        &+ \left[ E - V_c(z) - \frac{\hbar^2 p^2}{2 m(z)} \right] \bar{G} (\mathbf{p},z,z^\prime,E) = \delta(z - z^\prime) \: .
    \label{eq:Sch:Green:Fourier}
\end{align}
The LDOS is obtained from the imaginary part of the Green's function as~\cite{Akkermans_Montambaux:book}
\begin{equation}
    \mathrm{LDOS}(\mathbf{r},E) = - \frac{1}{\pi} \, \mathrm{Im} \, G(\mathbf{r}, \mathbf{r}, E) \: ,
\end{equation}
or in the $(x,y)$ Fourier space as
\begin{equation}
    \mathrm{LDOS}(\mathbf{p},z,E) = - \frac{1}{\pi} \, \mathrm{Im} \, \bar{G}(\mathbf{p}, z, z, E) \: .
\end{equation}
The latter expression measures the number of states with in-plane wave vector $\mathbf{p}$ and contributing at point $z$, per unit length and unit energy. Computing the LDOS thus requires solving Eq.~\eqref{eq:Sch:Green:Fourier} for the Green's function for a given energy $E$ and a given wave vector $\mathbf{p}$ (see Sec.~II in the SM for the numerical method). 

The calculated LDOS as a function of space and energy at vanishing in-plane wave vector ($p=0$) is shown in Figure~\ref{fig:DOS}(a). We observe that, at energies higher than the bulk CBM, the electronic wave functions are propagating plane waves in the vacuum and in the bulk of the semiconductor, as attested by the periodic spatial oscillations in the LDOS~\footnote{Note that we represent here the envelope function and not the full Bloch wave function in the semiconductor.}. For energies lower than the bulk CBM, electronic wave functions are also propagating plane waves in the vacuum and are evanescent waves in the bulk of the semiconductor. In the BBR, although there exists a continuum of states, the LDOS exhibits local maxima at specific energies, which we identify as resonant states. Their energies of about 2.4 and 3.0~eV above the Fermi level can be extracted from the LDOS integrated spatially over the semiconductor in Figure~\ref{fig:DOS}(b), and their dispersion as a function of the transverse component of the wavevector $p$ is plotted in Figure~\ref{fig:DOS}(c).
These states are metastable states in the BBR. They may be thought in a semiclassical picture as states for which electrons undergo many reflections within the BBR before being transmitted into the vacuum~\cite{Hatano:2008}. This is analogous to photons in a leaky Fabry-Perot cavity in optics, or more generally the so-called quasinormal modes in any structured open photonic system~\cite{Lalanne:2018,Both:2022}.
By fitting Lorentzian line shapes to the DOS, the lifetime of the two resonant states are estimated to be about 18 and 23~fs, for the lowest and highest energy resonant state, respectively.

Note that our numerical calculations in the open system framework remove restrictions from the early calculations~\cite{Korotkikh1978, Gerchikov1996}.  They show that the envelope wavefunction does not strictly vanish at the interface at resonances as can be seen in Figure~\ref{fig:DOS}(a), in contrast with the assumption usually encountered in the literature ~\cite{Korotkikh1978, Gerchikov1996}. 
Furthermore, the model shows that the thin barrier formed by the Cs layer is mostly transparent and that such resonant states exist even without it.
Analogous to modes in a Fabry-Pérot cavity, the resonance condition and the lifetime of the states depend both on the details of the potential on both sides of the interface and on the effective mass jump, as will be described in a future article. 
Perhaps more importantly, early calculations ~\cite{Korotkikh1978, Gerchikov1996, Orlov2000} did not address the spectral broadening or life-time of the resonant states which is fundamentally related to their metastability due to their leakage into vacuum. The Green's function approach in an open system presented here gives this information.

\begin{figure}[]
\includegraphics[width=0.75\columnwidth]{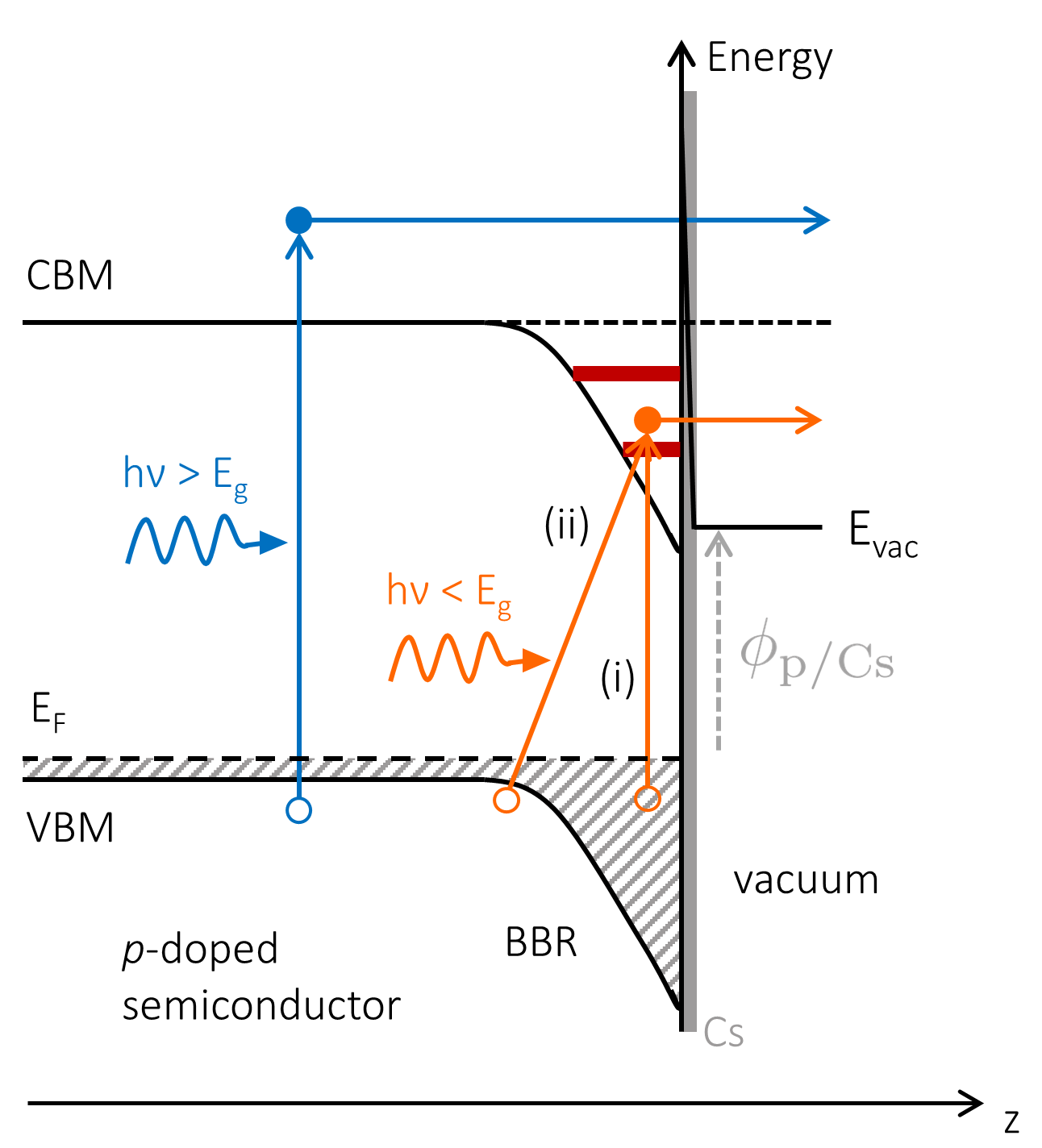}
\caption{\label{fig:schema} Schematic of the photoemission processes for a \textit{p}-type semiconductor in NEA, for near band gap excitation. The blue arrows schematize above band gap excitation, while the orange ones picture below band gap excitation and absorption in the BBR.}
\end{figure}

Experimentally, the existence of resonant states in the BBR of cesiated \textit{p}-GaN was investigated using low energy photoemission, as schematically shown in Figure~\ref{fig:schema}. The exposition of the sample to near band gap photons excites electrons in the semiconductor conduction band that can be emitted into the vacuum after transport and relaxation processes, in the bulk and in the BBR~\cite{Spicer1958, Scheer1965}. Emission to the vacuum is permitted by the low vacuum level due to the Cs monolayer at the sample surface \cite{Sauty-LEEM-Cs}.

For direct band gap semiconductors, with near but above band gap excitation, photoelectrons are generated in the first tens to first hundreds of nanometers of the material. Along their transport to the surface, they relax to the minima of the conduction band, which act as accumulation points and appear as characteristic features in the energy distribution of photoemitted electrons~\cite{Drouhin1985, Lassailly1990, Peretti1991, Piccardo2014, Sauty2022}, signature of the semiconductor DOS. 

On the contrary, with below band gap light excitation, band-to-band absorption in the bulk does not occur. A photoemission current can then arise only from absorption from defect states (process (i) in Figure~\ref{fig:schema})~\cite{Pakhnevich2004}, or from Franz-Keldysh absorption assisted by the strong electric field in the BBR (process (ii) in Figure~\ref{fig:schema}). After their excitation close to the surface, electrons do not experience any transport step. They are confined in the quasiwell formed by the BBR and can only relax in available states, before being emitted into vacuum. 

The studied sample is a MOCVD-grown c-plane heterostructure. It consists of 200-nm \textit{p}-GaN (Mg:$\SI{5e19}{\cm^{-3}}$) with surface overdoping (Mg:$\approx \SI{1e20}{\cm^{-3}}$ on the last 10 nm close to the surface), grown on \textit{n}-GaN (Si:$\SI{6e18}{\cm^{-3}}$) and UID GaN on a saphire substrate. The sample first underwent a chemical cleaning in piranha ($\mathrm{H_2SO_4}$ and $\mathrm{H_2O_2}$ mixture, 3:1 ratio) and HCl-isopropanol solutions~\cite{Tereshchenko2004}. After being inserted in the ultra high vacuum chamber (base pressure $\SI{5e-11}{\milli \bar}$), it was annealed for about 10 min at 350°C. To achieve NEA, cesium was deposited on its surface, while monitoring the photoemission current for an excitation energy corresponding to the semiconductor band gap (3.4~eV). A complete Cs monolayer was obtained~\cite{Sauty-LEEM-Cs}, and was stable for at least one day, during which data were acquired, at room temperature.

The energy distribution of the photoemitted electrons was then recorded using a UV-enhanced lamp associated with bandpass filters, with a bandwidth of 10~nm, for an excitation power of about \SI{200}{\micro\W}200~µW. The electrons were collected by an energy analyzer specifically designed for low energy electrons~\cite{Drouhin1986}, with an energy resolution of 50~meV. Derivatives of the energy distribution curves were obtained numerically, and the resolution was decreased to 80~meV due to a smoothing step in the data processing. The experimental setup is identical to previously published work ~\cite{Peretti1991, Piccardo2014, Sauty2022}.

\begin{figure}[h!]
\includegraphics[width=1.0\columnwidth]{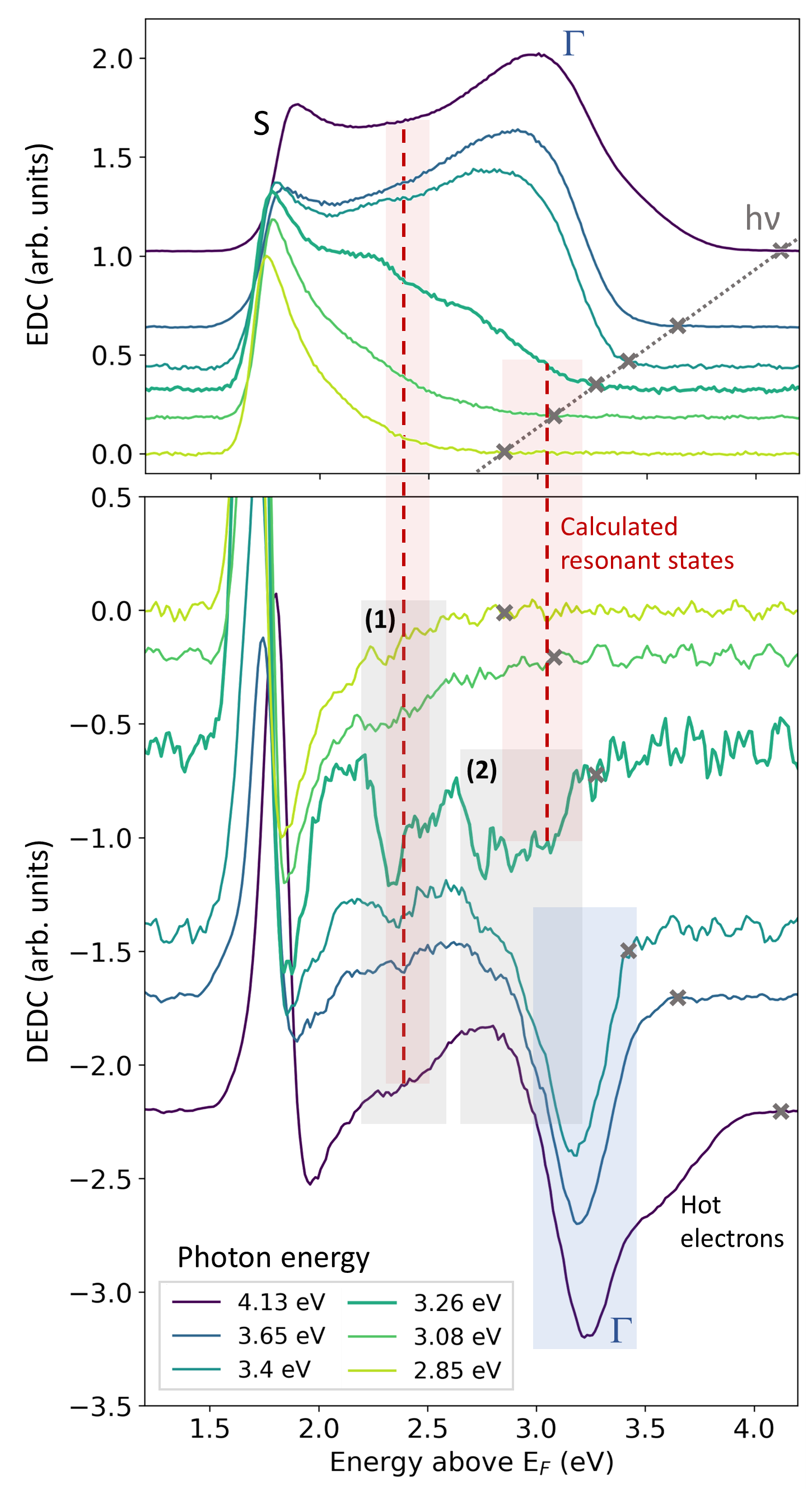}
\caption{\label{fig:EDC} EDCs and DEDCs on GaN for near band gap excitation, from 2.85 to 4.13~eV. The vacuum level is at about 1.5~eV above the Fermi level ($E_\mathrm{F}$), meaning that NEA is achieved. For photon energies $h\nu >$ 3.4~eV, the $\Gamma$ contribution from electrons accumulated in the bulk CBM, appears at identical position in all curves (blue shaded area). Additional contributions at lower energy than $\Gamma$ are marked by the two gray-shaded area in the DEDC. The positions of the calculated resonant states are indicated by dashed red lines.}
\end{figure}

The energy distribution curves (EDCs) obtained for excitation energies between 2.85 and 4.13~eV are shown in Figure~\ref{fig:EDC}, along with their numerical derivatives (DEDCs). Since the GaN band gap is 3.4~eV, half of the curves correspond to below band gap excitation. For clarity, both EDC and DEDC have been translated along the \textit{Y} axis, the EDC translation being proportional to the incident photon energy. 
For all curves, the vacuum level, which corresponds to the low-energy threshold of the EDC, is at about 1.5~eV above the Fermi level, confirming that surface NEA is achieved. The high-energy threshold of the EDC corresponds to the energy $E_F + h\nu$ (marked in dashed line on the EDC plot).

Depending on excitation energy, different contributions appear. For all excitation energies, the lowest energy contribution, marked S in the EDC, corresponds to the continuum of electrons excited close to the surface~\cite{Piccardo2014} (process shown by orange arrows in Figure~\ref{fig:schema}). In addition, for above band gap excitation ($h\nu >$ 3.4~eV), the contribution from electrons accumulated in the bulk CBM, noted $\Gamma$, appears as a positive bump in the EDC, whose high-energy side forms a negative feature in the DEDC, shaded in blue. This $\Gamma$ contribution has a high-energy threshold at the bulk CBM position, and extends to lower energy due to the electron energy losses when crossing the BBR~\cite{Piccardo2014}. 
Between these two contributions (S and $\Gamma$), two other peaks appear progressively when the excitation energy is increased. They are shaded in gray in the DEDC and labeled (1) and (2). They are particularly enhanced for an excitation energy of 3.26~eV, for which they are both accessible energetically by electrons but are not yet hidden by the $\Gamma$ contribution. 
Note that contribution (1) appears for all excitation energies although nearly hidden in the noise level in some of the curves. Contribution (2) is visible only for excitation energies above 3.26 eV and for excitation energies above 3.4~eV it appears as a low-energy shoulder of the $\Gamma$ contribution.
We attribute these two contributions (1) and (2) to the resonant states in the open well formed by the BBR and the sample surface.

The energy position of the resonant states in the BBR obtained by our LDOS calculation for a surface overdoping of $\SI{1e20}{\cm^{-3}}$ are reported as dashed red lines on the experimental curves in Figure~\ref{fig:EDC}, at 2.4 and 3.0~eV above the Fermi level. The red shaded area around the calculated resonant states corresponds to energy positions obtained for doping levels between $\SI{5e19}{\cm^{-3}}$ and $\SI{2e20}{\cm^{-3}}$, reflecting the uncertainty on the exact surface doping level. Overall, the calculated energy positions match the two contributions observed in the DEDCs. 
To compare more precisely the experimental contributions to the calculation, one must consider the additional sources of broadening apart from the intrinsic lifetimes of the states. Typically, we can expect that \textit{p}-doping heterogeneities as well as the resonant states dispersion with the wave vector shown in Figure~\ref{fig:DOS}(c) might induce energy broadening but more importantly, the shape of the contribution of a resonant state in the DEDC will result from the interplay between the escape of the electrons accumulated in the state and their relaxation toward the continuum. In GaN, the optical phonon scattering rate is of the order of 10-30 fs \cite{Marcinkevicius2016, Tsen1997, Bertazzi2009}. This is comparable with the lifetimes of the calculated resonant states of 18 and 23~fs and explains that in the experimental curves, the states contributions extend to lower energies than the calculated resonance energies.

Finally, these BBR resonant states are very clearly distinguishable only for specific excitation energies below the material band gap, for which they can be populated but their contribution is not hidden by the numerous energy-relaxed  bulk electrons. In the previous attempts to observe such states in GaAs photocathodes \cite{Orlov2000, Korotkikh1978, Jin2015}, such a condition was harder to obtain due to the smaller band gap of GaAs, and we believe that in the published experimental data, the resonant states contribution could not be distinguished from the bulk $\Gamma$ contribution, as we explain in more details of Sec.~III of the SM.

In this work, we have investigated, both experimentally and theoretically, the DOS in the surface BBR of cesiated p-type GaN. We have shown that despite the opening of the potential well by the reduction of the workfunction by the Cs layer, resonant metastable states exist in the energy continuum extending from the vacuum level to the CBM in the bulk.
On the experimental side, the combination of a below band gap excitation and a large band gap material permits the observation of such states without them being hidden by the large contribution of bulk electrons. On the theory side, our Green's function approach allows the calculation of the states lifetimes.

We thank Lucia Reining and Marcel Filoche for fruitful discussions. 
This work was supported by the French National Research Agency (TECCLON Grant No. ANR-20-CE05-0037-01, CPJ LUMIS Grant No. ANR-25-CPJ1-0059-01) and by the Simons Foundation (Grants No. 1027114 C.W., No. 601944 and 601937 J.-P. B.).
Support at UCSB was provided by the Solid State Lighting and Energy Electronics Center (SSLEEC); U.S. Department of Energy under the Office of Energy Efficiency \& Renewable Energy (EERE) Award No. DE-EE0009691; the National Science Foundation (NSF) RAISE program (Grant No. 
DMS-1839077); the Simons Foundation (Grant No. 601952). 
\medskip

Additional references cited in the Supplemental Material: \cite{Weisbuch:book, Bastard:1981,Levy-Leblond:1995,Einevoll:1994,BornWolf,singh_2003,nr}

\bibliography{biblio}

\end{document}